\documentclass[useAMS,usenatbib,a4]{mn2e}
\usepackage{graphicx}% This allows you to call \includegraphics
\usepackage{color}
\def\etal{{\it et al}}

\def\P3hat{{\mathaccent 94 P}_3}

\def\eg{{\it e.g.}}

\def\aap{A\&A}

\def\apj{Ap.J.}
\def\apjs{Ap.J. Suppl.}
\def\mnras{M.N.R.A.S.}
\def\nat{Nature}

\newcount\notenumber
\def\clearnotenumber{\notenumber=0}
\def\note{\advance\notenumber by1 \footnote{$^{\the\notenumber}$}}
\clearnotenumber
\usepackage{rotating}   % to rotate the expressions and tables.

\title[Modulation, Polarization \& Carousel of B1857--26]
{On the Subpulse Modulation, Polarization and Subbeam Carousel Configuration of Pulsar B1857--26}

\author[Dipanjan Mitra \& Joanna Rankin] 
{Dipanjan Mitra$^{1}$ \& Joanna M. Rankin$^{2}$ \\ 
$^1$National Centre for Radio Astrophysics, Ganeshkhind, Pune 411 007 India : dmitra@ncra.tifr.res.in\\
$^2$Physics Department, University of Vermont, Burlington, VT 05405 USA : Joanna.Rankin@uvm.edu \\
}

\date{Released 2004 Xxxxx XX}

\pagerange{\pageref{firstpage}--\pageref{lastpage}} \pubyear{2004}

\def\LaTeX{L\kern-.36em\raise.3ex\hbox{a}\kern-.15em
    T\kern-.1667em\lower.7ex\hbox{E}\kern-.125emX}

\begin{document}

\label{firstpage}

\maketitle

\begin{abstract}
New GMRT observations of the five-component pulsar B1857--26 provide 
detailed insight into its pulse-sequence modulation phenomena for the 
first time.  The outer conal components exhibit a 7.4-rotation-period, 
longitude-stationary modulation.  Several lines of evidence indicate a carousel 
circulation time $\P3hat$ of about 147 stellar rotations, characteristic of a 
pattern with 20 beamlets.  The pulsar nulls some 20\% of the time, usually 
for only a single pulse, and these nulls show no discernible order or 
periodicity.  Finally, the pulsar's polarization-angle traverse raises interesting 
issues:  if most of its emission is comprised of a single polarization mode, the 
full traverse exceeds 180\degr; or if both polarization modes are 
present, then the leading and the trailing halves of the profiles exhibit two 
different modes.  In either case the rotating vector model fails to fit the 
polarization-angle traverse of the core component.
\end{abstract}


\section*{I. Introduction} 
Pulsar B1857--26, though discovered early (Vaughn \& Large 1970), has heretofore 
been studied almost entirely by average methods.  One of the small group of pulsars 
with five-component ({\bf M}) profiles, its aggregate polarization has been investigated 
over a broad frequency band (Hamilton \etal\ 1977; McCulloch \etal\ 1978; Manchester 
\etal\ 1980; Morris \etal\ 1980; van Ommen \etal\ 1997; \& Gould \& Lyne 1998), and no 
effort to interpret the form of pulsar beams fails to mention it (Backer 1976; Lyne \& 
Manchester 1988; Rankin 1983a, 1986, 1990, 1993a,b; Mitra \& Deshpande 1999).  
Little has been learned, however, regarding its pulse-sequence (hereafter PS) 
behaviour.  Apart from one historical effort to determine its null fraction (Ritchings 1976), 
only recently did the fluctuation-spectral analyses of Weltevrede \etal\ (2006, 2007; 
hereafter WES, WSE) identify its 7-rotation-period (hereafter $P_1$) modulation.  
PSR B1857--26 is often compared with other prominent {\bf M} pulsars, but it is not yet 
known whether it exhibits profile moding like B1237+25 (\eg, Srostlik \& Rankin 2005) 
or core-component phenomena such as those seen in B0329+54 (Mitra \etal\ 2007).  

\begin{figure} 
\begin{center}
\includegraphics[width=72mm,angle=-90.]{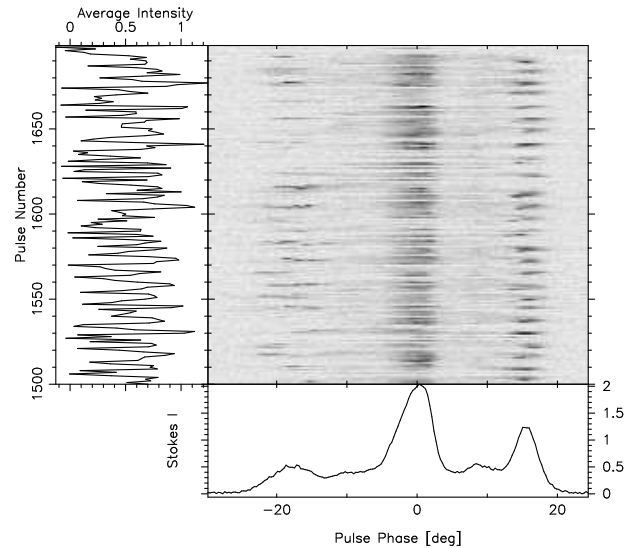}
\caption{A 200-pulse Stokes-$I$ pulse sequence from pulsar B1857--26 (central 
panel).  The average profile and integrated intensity are given in the bottom and 
side panels, respectively.  Note that the trailing component exhibits a regular 
7-rotation-period, longitude-stationary modulation that is also perceptible in the 
leading component.  The intensity scale in this and subsequent figures in 
arbitrary.}
\label{fig1}
\end{center}
\end{figure}

We have carried out a sensitive new 325-MHz polarimetric observation of B1857--26 
using the Giant Metre-Wave Radio Telescope (hereafter GMRT) in Maharashtra, and 
a typical 200-pulse segment is given in Figure~\ref{fig1}.  Already we see evidence 
of interesting subpulse modulation.  In the remainder of this paper, we describe the 
results of a series of PS analyses designed to address a number of these questions 
remaining about this star's individual pulse behaviour.  To our surprise we found 
that we were able to identify and measure a long period cycle which almost certainly 
can be interpreted as the rotation interval of an emission-beam carousel in the manner 
of that found earlier for B0943+10 (Deshpande \& Rankin 1999, 2001; hereafter DR99, 
DR01).  Such cycles are thought to be driven by ${\bf E}$$\times$${\bf B}$ forces 
within a pulsar's polar flux tube along the lines of the Ruderman \& Sutherland 
(1975; hereafter R\&S) theory; however, all those so far determined or reliably 
estimated\footnote{B0834+06: Asgekar \& Deshpande 2005; Rankin \& Wright 2007a; 
B0809+74: van Leeuwen \etal\ 2003; B0834--26: Gupta \etal\ 2004; B1133+16: 
Herfindal \& Rankin 2007; J1819+1305: Rankin \& Wright 2007b.} are longer or 
much longer than predicted by the theory.  \S II describes the GMRT observations, 
and \S III our efforts to measure and model its polarisation-angle traverse, so as 
to better determine the emission geometry.  In \S IV we analyse the properties 
of the nulls, and in \S V we outline the dynamics of its core component. \S VI gives a 
discussion the fluctuation-spectral analyses, and \S VII the evidence for a tertiary 
modulation cycle.   \S VIII then discusses the configuration of the pulsar's subbeam 
carousel, and \S IX gives a brief discussion and summary of the results.

\section*{II. Observations} 
\label{sec2}

Pulse-sequence polarisation observations of pulsar B1857--26 were acquired using 
the Giant Meterwave Radio Telescope (GMRT) north of Pune, India at 325 MHz on 
2004 August 27.  The GMRT is a multi-element aperture-synthesis telescope consisting 
of 30 antennas which can be configured as a single dish. The polarimetry discussed 
here combined the array signals coherently in the upper of the two 16-MHz `sidebands' 
in a manner identical to that described for pulsar B0329+54 in Mitra \etal\ (2007).  The 
pulsar back-end computed both the auto- and cross-polarized power levels, which were 
then recorded at a sampling interval of 0.512 msec.  A suitable calibration procedure 
as described in Mitra \etal\ (2005) was applied to the recorded observations to recover 
the calibrated Stokes parameters $I$, $Q$, $U$ and $V$.  The duration of the PS is 19.4 
minutes or 1945 pulses.  In addition, a total-power observation at 610 MHz with the same 
bandwidth, sampling time and duration was carried out on 2007 February 4.  

\begin{figure}
\begin{center}
\includegraphics[width=80mm,angle=-90.]{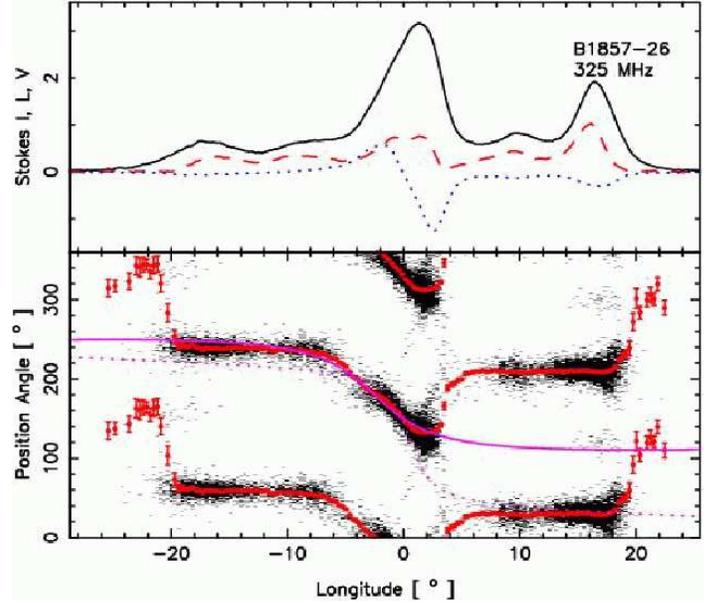}
\caption{PA histograms for the entire 325-MHz PS.  The upper panels 
show the aggregate total power, total linear and circular polarisation
(LH-RH), and the lower panel gives the polarization-angle (hereafter 
PA) density.  The PA values are plotted twice for clarity, and the two 
curves indicate negative-going RVM fits to the PA traverse.  One 
RVM fit (dotted curve) assumes that both sides of the profile represent 
the PPM and, clearly, it fails to describe the $>$180\degr extent of this 
apparent PA rotation.  A second RVM fit assumes that different OPMs 
predominate in the leading and trailing parts of the profile.  The longitude 
origin is taken near the zero-crossing point of the antisymmetric Stokes 
$V$ signature.}
\label{fig2}
\end{center}
\end{figure}

\section*{III. Emission Geometry}
\label{sec3} 
A histogram of the polarisation-angle (hereafter PA) behaviour of the 
325-MHz observation is given in Figure~\ref{fig2}.  The upper panel 
shows  the star's five components in total intensity (Stokes $I$), the 
total linear polarisation (Stokes $L$ [=$\sqrt{Q^2+U^2}$]) and the 
circular polarization (Stokes $V$ [=LH--RH]); whereas the lower 
panel gives the PA density twice for clarity.  Several aspects of this 
diagram deserve close discussion.  

First, the very flat PA curves under the wings of the profile appear 
compatible with a positive (equatorward) sense of the sightline 
impact angle $\beta$ as our attempts to fit a rotating-vector model 
(hereafter RVM: Radhakrishnan \& Cooke 1969; Komesaroff 1970) 
to the traverse bear out.  Gould \& Lyne's (1998) polarimetry confirms 
this aspect of the pulsar's PA traverse at meter wavelengths.  At 1.4 
GHz, however, the traverse under the wings of the profile is not at all 
flat and even shows different slopes on the leading and trailing 
sides of the profile (see also Johnston \etal\ 2005).  Clearly, such 
behaviour is difficult or impossible to reconcile with the RVM.

Second, no sensible RVM fit\footnote{The RVM fits done here use the method and
convention described by Everett \& Weisberg (2001). Here, the goodness of the
fits is assessed by the reduced chi-square, which in the present case
yielded significantly large values. Also $\alpha$ and $\beta$ are correlated
up to 98\%.} could be obtained to the PA traverse 
under the assumption that most of the star's emission, throughout 
the profile, stems from a single orthogonal polarisation mode 
(hereafter OPM)---here the putative primary polarization mode 
(hereafter PPM).  All the published observations seem to confirm 
that the PA traverse rotates negatively (clockwise) in the center of 
the profile, but if most of the emission reflects the PPM, then the 
full PA traverse significantly exceeds 180\degr\ both at meter and 
centimeter wavelengths.  While the RVM PA excursion can 
exceed 180\degr for inner (poleward) sightline traverses, it cannot 
do so for outer ones, thus some different interpretation is needed.  
This difficulty is illustrated by the fit (dotted curve) in the bottom 
panel of Fig.~\ref{fig2}.

\begin{figure} 
\begin{center}
\includegraphics[width=78mm,height=84mm,angle=-90.]{PQB1857-26_325_lrf_1-1900.ps}
\includegraphics[width=78mm,height=84mm,angle=-90.]{PQB1857-26_lrf_610_1-600.ps}
\end{center}
\end{figure}
\begin{figure} 
\begin{center}
\includegraphics[width=78mm,height=84mm,angle=-90.]{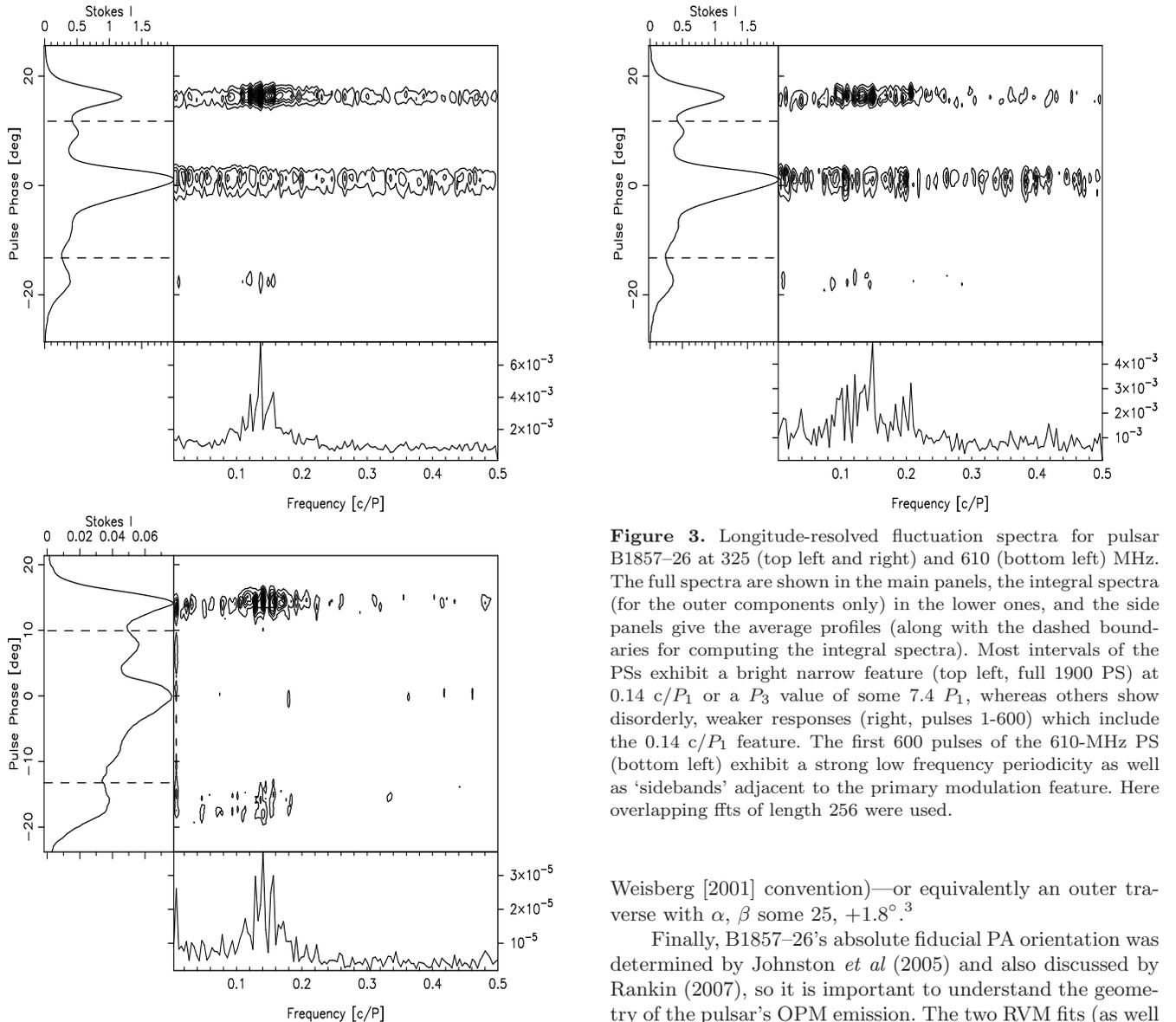}
\caption{Longitude-resolved fluctuation spectra for pulsar B1857--26 
at 325 (top left and right) and 610 (bottom left) MHz.  The full spectra are 
shown in the main panels, the integral spectra (for the outer components only) 
in the lower ones, and the 
side panels give the average profiles (along with the dashed 
boundaries for computing the integral spectra).  
Most intervals of the PSs exhibit a 
bright narrow feature (top left, full 1900 PS) at 0.14 c/$P_1$ or a $P_3$ 
value of some 7.4 $P_1$, whereas others show disorderly, weaker 
responses (right, pulses 1-600) which include the 0.14 c/$P_1$ feature.  
The first 600 pulses of the 610-MHz PS (bottom left) exhibit a strong low 
frequency periodicity as well as `sidebands' adjacent to the primary 
modulation feature.  Here overlapping ffts of length 256 were used.}
\label{fig3}
\end{center}
\end{figure}

We also tried to fit B1857--26's PA trajectory making the somewhat 
unsavory assumption that different OPMs are dominant in the leading 
and trailing wings of the pulse profile (as we know of no excellent 
example of this configuration in other pulsars).  This RVM fit is indicated 
by the thinner full curve in Fig.~\ref{fig2}, and note that it is hardly 
excellent.  It results in a large (some 3\degr) displacement between 
the PA inflection point and the zero-crossing longitude of the circular 
polarization. 

One characteristic of the profile possibly supporting this interpretation 
is the deep minimum in the total linear polarization just following the 
core component, which is prominent in every published observation.  
This latter fit results in values for the magnetic latitude $\alpha$ and 
sightline-impact angle $\beta$ of 155 and --1.8\degr, respectively 
(using the Everett \& Weisberg [2001] convention)---or equivalently an 
outer traverse with $\alpha$, $\beta$ some 25, +1.8\degr.\footnote{Even 
had these fits been fully successful, they would not close all 
questions about the pulsar's PA behaviour.  We also see a weak track 
under the core component suggesting a positive PA traverse (where 
the average PA tends to follow on the core's trailing edge) with a slope 
of perhaps +40-50\degr/\degr.  Recent studies (\eg, Srostlik \& Rankin 
2005; Mitra \etal\ 2007) show that core linear polarisation does not 
always follow the RVM, and here it is concentrated in two ``spots'' near 
the +/-- circular maxima, rather than indicating any clear RVM track.  
Were this the correct {\it geometrical} interpretation, $\alpha$, $\beta$ 
would be some 27, +0.6\degr.}

Finally, B1857--26's absolute fiducial PA orientation was determined 
by Johnston \etal\ (2005) and also discussed by Rankin (2007), so it 
is important to understand the geometry of the pulsar's OPM emission.  
The two RVM fits (as well as the further conjecture) above make 
very different assumptions about this geometry, and they result in different 
offsets between the centres of the linear and circular polarisation.  Further 
study at multiple frequencies is needed to resolve these issues;
but, fortunately, we need not settle them fully for our present purpose.

\section*{IV.  Nulling Behaviour} 
\label{sec4} 
B1857--26 exhibits some 20$\pm$3\% null pulses, substantially more 
than the 10$\pm$2.5\% indicated by Ritchings' (1976) very old analysis.  
We computed null histograms for the observations similar to Redman 
\etal's fig. 12.  In both, the pulsar maintains a relatively constant brightness 
(so scintillation effects are minimal) and the noise level is such that about 
0.3 $<$$I$$>$ provides a plausible threshold for null identification.  The 
intensity distributions of the pulsar signal and noise overlap so that weak 
pulses cannot be reliably distinguished from nulls.  Nonetheless, we see 
that the pulsar's nulls are very short, with a strong propensity for 1 $P_1$ 
duration, and a maximum length in our observations of 4 $P_1$.  Its bursts 
are also short, many having only a single period, and a mean burst length 
of only 5 $P_1$.  We find little evidence that the nulls are non-random or 
even weakly periodic (using the methods of Herfindal \& Rankin 2007).  
However, the average profile of the putative nulls (not shown) does 
indicate significant power at the longitude of the central component,  
whereas the partial profiles of pulses just before and just after nulls show 
no significant difference in form from the total average profile.

\begin{figure} 
\begin{center}
\includegraphics[width=92mm,height=84mm,angle=-90.]{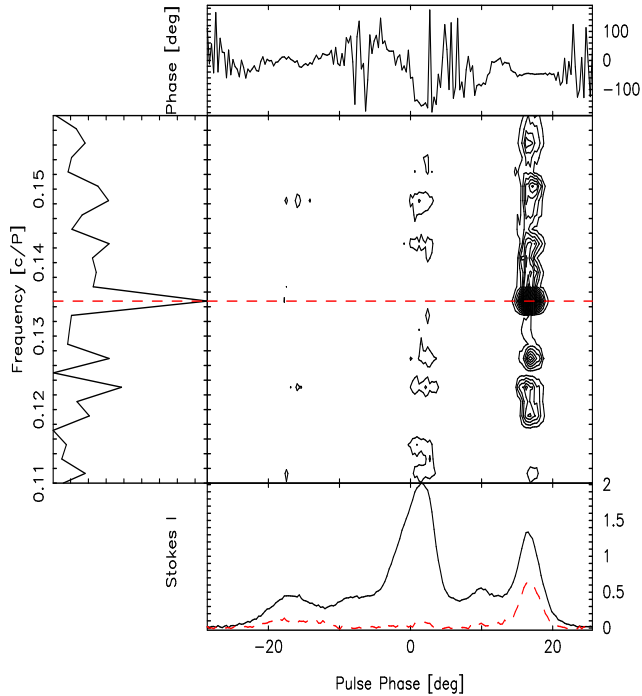}
\caption{Modulation amplitude (bottom panel dashed curve) and phase (top panel) 
for the pulse 1101-1612 sub-PS in Fig.~\ref{fig3} (top left), shown as a 
function of the average-profile power (solid curve bottom panel). The 
central panel shows the contour levels for the lrf (zoomed in between 
0.08 to 0.2 c/$P_1$) and the left panel shows the longitude-averaged 
fluctuation spectrum.  Note that virtually all of the fluctuating power here 
falls under the outer conal component pair (I \& V)---and primarily V.} 
\label{fig4}
\end{center}
\end{figure}

\section*{V. Core Dynamics}
\label{sec5}
The PS properties of core-component emission have not been well 
studied, and B1857--26 provides a further important opportunity to examine 
its characteristics.  This pulsar's core tends to be quite regular in intensity, 
as is suggested by Fig.~\ref{fig1} and further indicated by its rather small 
modulation indices at both 92 and 21 cms (WES, WSE).  The core feature 
is not at all symmetrical (or of Gaussian form) with its slow rise and steeper 
fall, and its constituent subpulses exhibit neither the intensity-dependent 
form nor the longitude shift seen in pulsar B0329+54 (\eg, Mitra \etal\ 2007).  
Even its antisymmetric circular polarisation is rather consistent, showing very 
nearly the same phase in every pulse.  Interestingly, the strongest single 
pulses seem always to fall near the core peak and are highly left-hand 
circularly polarised; whereas the leading right-hand circular is somewhat 
weaker and steadier (see also WSE's asymmetric variance profile).  Also, 
we see a slight change in the total linear polarisation from a double form 
at low intensities to an emphasis on the trailing lobe in the strongest pulses.  
Again, the circularly polarised core signature in B1857--26 bears careful 
examination as it fails to be fully antisymmetric in just the same manner as 
the Stokes $I$ profile fails to be symmetric.  Nonetheless, its zero-crossing 
point falls very close to the center of the conal profile.

\section*{VI. Fluctuation-Spectral Analyses}
\label{sec6}
Figure~\ref{fig3} gives longitude-resolved fluctuation (hereafter lrf) spectra 
for three different segments of the two PSs. 
As expected from Fig.~\ref{fig1} 
periodic modulation is seen in the outer conal components (I and V, and its contribution
is shown in the integrated fluctuation spectra in the bottom panels) which 
often exhibits a strong feature near 0.14 cycle/$P_1$ (hereafter c/$P_1$).  The 325-MHz lrf (top 
left) shows that this narrow feature dominates throughout the entire PS.  
Atop the broader response, its power falls in a single bin of a 512-length 
FFT (not shown) with a frequency of 0.135$\pm$0.001 c/$P_1$ or 
7.41$\pm$0.06 $P_1$.  The corresponding harmonic-resolved fluctuation 
spectrum (hereafter hrf, see DR01 for details; also not shown) indicates 
that the respective positive and negative responses represent mostly 
amplitude modulation as expected.  Figure~\ref{fig4} shows the modulation 
amplitude and phase for the interval 1101--1612 PS.  Virtually all the power with 
this periodicity falls under the outer conal components (not the inner pair), 
and the flat phase gradient in this region indicates a longitude-stationary 
modulation---just as expected for a highly central sightline traverse.

Other sections of the 325-MHz PS, however, show less regularity. The 
right-hand lrf plot of Fig.~\ref{fig3} gives a representative example.  Here, 
the fluctuations produce several broad features around the primary one.  
We emphasize that the 7-$P_1$ modulation is associated with the outer conal 
components and primarily the trailing one;  it cannot even be {\it detected} 
in the inner conal region of the profile.  The fluctuation spectra of the 
outer components is conflated in the integral spectra, but separate analyses 
(not shown) of the conal and core parts of the profile show that the core 
contributes little (even noise power) to the integral spectra.  In summary, 
only the outer cone fluctuates significantly at the roughly 7-$P_1$ cycle. 

The bottom lrf of Fig.~\ref{fig3} was computed from the first 600 pulses of 
the 610-MHz observation (using overlapping 256-length ffts).  It 
shows both a low frequency feature and two `sidelobes' adjacent to the 
primary modulation feature, possibly indicating a tertiary modulation 
as in B0943+10 (see DR01).  The primary modulation has a period of some 
7.11$\pm$0.01 $P_1$, whereas the entire PS gives 7.34$\pm$0.01 $P_1$,  
suggesting that the primary modulation frequency is somewhat variable.  
The low frequency feature can be measured over an interval of up to 
1000 pulses, and the hrf (not shown) indicates that it represents a mixture 
of amplitude and phase modulation with a period of about 145 $P_1$.  
The two `sidebands' are not equally spaced, but the average interval is 
about 0.013 c/$P_1$.  

\begin{figure}
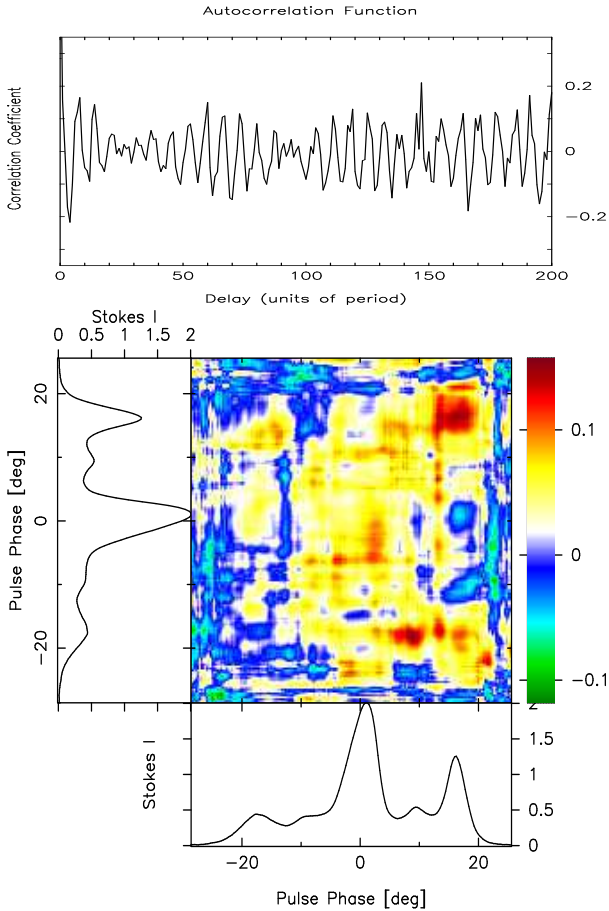
 
\begin{center}
\includegraphics[width=40mm,height=80mm,angle=-90.]{b1857-26_325_800-1100_ACF.ps}
\includegraphics[width=80mm,height=80mm,angle=-90.]{PQB1857-26_325_lc_800-1700_delay147.ps}
\caption{The top plot shows the correlation coefficient of pulses 
800-1100 of the 325-MHz PS in Figs.~\ref{fig3} \& \ref{fig4} using only the 
trailing longitude regions. The bottom display gives the longitude-longitude 
correlation for a delay of 147 $P_1$ over a longer PS of 800-1700.  The 
main panel gives the correlation according to the color bar on the right, 
and the average profiles are plotted on both axes for reference.  The 
ordinate has been delayed with respect to the abscissa. The 7-$P_1$ 
cycle is clearly seen in top plot.  Note, however, the extraordinary peak 
at 147 $P_1$ which shows some 30\% correlation in the top plot and is 
clearly seen in the trailing component in the bottom display.}
\label{fig5}
\end{center}
\end{figure}

\begin{figure}
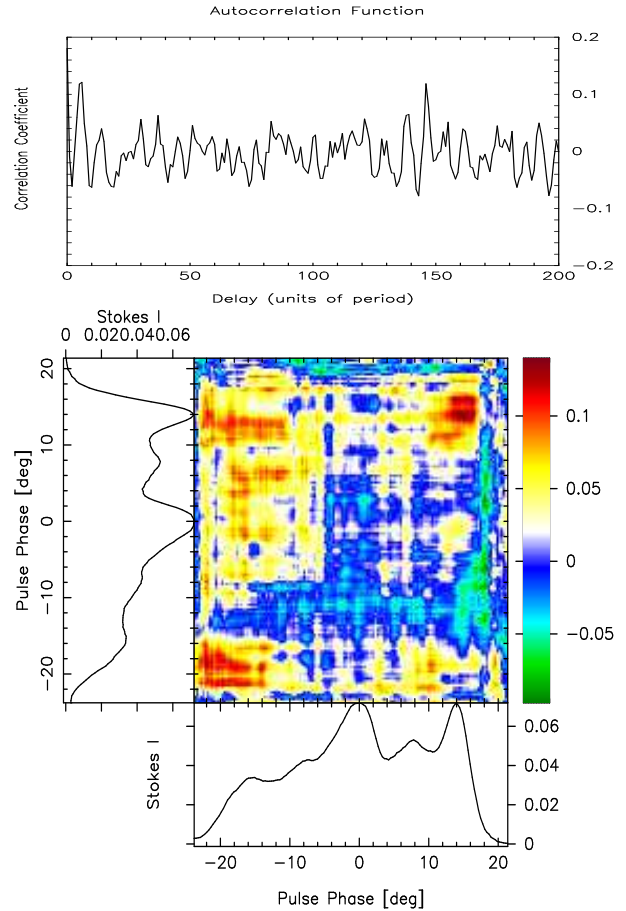

\begin{center}
\includegraphics[width=40mm,height=80mm,angle=-90.]{B1857-26_610_AUTOCORR_800-1900.ps}
\includegraphics[width=80mm,height=80mm,angle=-90.]{PQB1857-26_610_lc_800-1900_delay147.ps}
\caption{ Correlation plots as in Fig.~\ref{fig5} of pulses 800-1900 
of the 610-MHz PS using both the leading and trailing components.  
Note the peak at 147 $P_1$.  High correlation regions associated 
with both of these components are clearly visible in the bottom display 
at a delay of 147 $P_1$ (see text for details).}
\label{fig5a}
\end{center}
\end{figure}

\begin{figure}
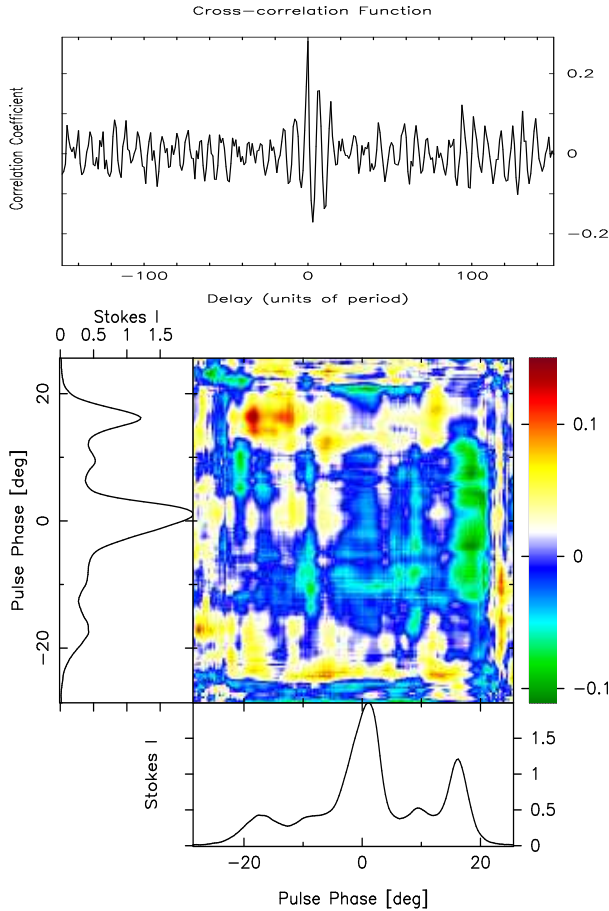
 
\begin{center}
\includegraphics[width=40mm,height=80mm,angle=-90.]{npdelay_325.ps}
\includegraphics[width=80mm,height=80mm,angle=-90.]{PQB1857-26_325_lc_800-1944_delay94.ps}
\caption{The top plot shows the cross-correlation between components I and 
delayed V of pulses 800-1944 of the 325-MHz PS.  Note the correlation peaks 
around +94-95 $P_1$ of up to 12\%, significantly higher than in the adjacent regions; 
however, no extraordinary correlation is seen at a lag of --53 $P_1$.  The +94-$P_1$ 
delay longitude-longitude correlation is shown in the lower display, where the correlation 
between components I and delayed V is evident. 
}
\label{fig5b}
\end{center}
\end{figure}

\begin{figure}
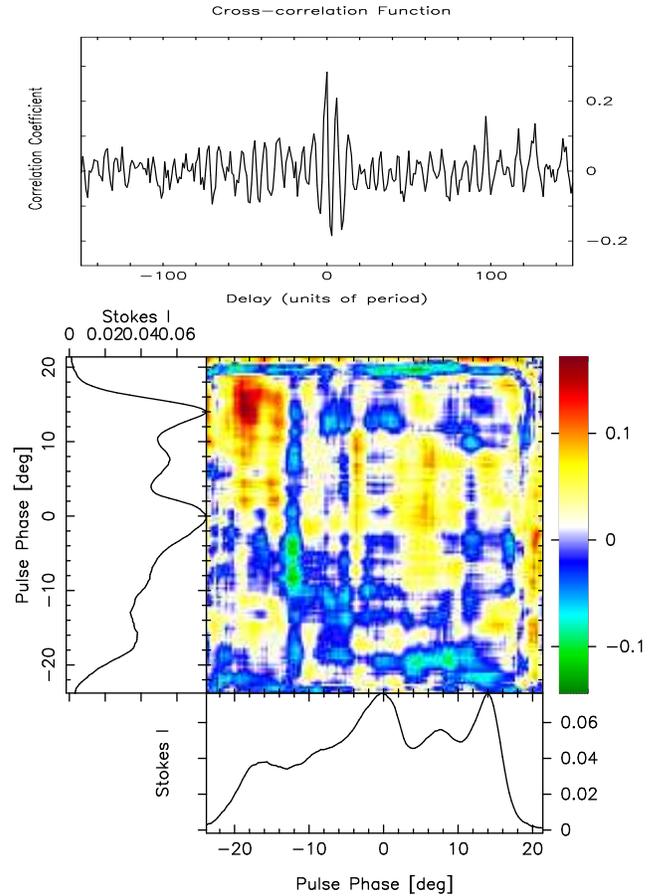
 
\begin{center}
\includegraphics[width=40mm,height=80mm,angle=-90.]{npdelay_610.ps}
\includegraphics[width=78mm,angle=-90.]{PQB1857-26_610_lc_1-600_delay97.ps}
\caption{Cross-correlation plots as in Fig.~\ref{fig5b} between components I 
and delayed V of pulses 1-600 of the 610-MHz PS.  Again we see a 
prominent peak at a lag of +97 $P_1$, but little at --50 $P_1$.  The bottom 
panel shows the longitude-longitude correlation map of the same 610-MHz 
PS interval in Fig.~\ref{fig3} (bottom) at a delay of +97 $P_1$.  Here the PS 
has been smoothed by 5 samples.  Note the strong correlation between 
component I and the delayed component V.}
\label{fig6}
\end{center}
\end{figure}

\section*{VII.  Evidence of a Tertiary Modulation Cycle}
\label{sec7}
We now enquire whether the 7-$P_1$ cycle is a part of a longer tertiary 
modulation cycle as might be expected if the observed modulation were 
produced by a rotating-subbeam ``carousel'' as found, in particular, for 
B0943+10 (DR99, DR01).  The long period lrf feature in the 610-MHz PS 
(Fig.~\ref{fig3}, bottom) is suggestive of such a cycle---as are its pair 
of sidebands. It is interesting to note that such a feature was also found 
in recent Westerbork observations at both 92 (WSE) and 21 cms 
(WSE)---but its period there cannot be determined accurately.  However, 
an autocorrelation-function (ACF) analysis of several sections of the the 
325-MHz, pulse 800-1944 interval identifies a cycle of similar length. The 
top panel of Figure~\ref{fig5} gives the result of this analysis done for the 
outer trailing conal component of the 800-1100 $P_1$, where the 7-$P_1$ 
corrugation is prominent as expected.  Note the prominent peak at a lag 
of 147 $P_1$, which is close to 20 times $P_3$.  Indeed, this added 
positive correlation peak is superposed on the 20th 7-$P_1$ corrugation.  
This peak has an amplitude of about 30\%, which is 8 times the expected 
error in the ACF.  It is interesting to note the structure in the 7-$P_1$ 
corrugation, which also has a periodicity of about 147 $P_1$.  The ACF 
seems to be consistent with a functional form of the type $\cos( 2 \pi d /147) \cos(2 \pi d /7)$, 
where $d$ is the delay in units of $P_1$.  Such orderly variations in the 
correlation could reflect non-uniform patterns in either the spacing or 
amplitude of the carousal beamlets.
The longitude-longitude correlation at a lag of 147 $P_1$ (for a slightly 
longer PS of 800-1700) is displayed in the lower part of Fig.~\ref{fig5}. 
This sequence exhibits slightly lower (yet significant) correlation in the 
strong trailing conal component.  The ACF analysis of the 610-MHz 
observation also shows significant correlation at 146 $P_1$ with a 
peak at 147 $P_1$ as illustrated in the top panel of Figure~\ref{fig5a}.  
Unlike the 325-MHz PS, here significant correlation at 147 $P_1$ is 
seen for both the leading and the trailing components as patches of 
red along the diagonal in the longitude-longitude correlation map 
(bottom display).

If this 147-$P_1$ peak is indicative of the tertiary modulation, which might 
be produced by a subbeam carousel system, then we might expect that 
there should be appreciable correlation between components I and V at 
some part of the 147-$P_1$ cycle, reflecting the circumstance that the 
`beamlets' rotate around the outer cone.  Consideration of the emission 
geometry (see \S III and Fig.~\ref{fig7} below) suggests that this interval 
should be a about 1/3 of the cycle.  In both the 325-MHz (pulses 800-1944) 
and 610-MHz PSs (over pulses 1-600; see Fig.~\ref{fig3} bottom), we find 
significant cross-correlation between comp. I and delayed comp. V at  
lags of 94-95 and 96-97 $P_1$, respectively, as shown in Figs.~\ref{fig5b} 
and \ref{fig6}.   We find no extraordinary cross-correlation between 
comp. V and delayed comp. I, which we had expected at lags of 53-54 
$P_1$ at 325 MHz (see Fig.~\ref{fig5b}) and 50-51 $P_1$ at 610 MHz 
(if we assume the exact circulation time is 147-$P_1$).  This might result 
if the actual peak lag fall between two integers.  These various correlations 
reiterate that the carousel `beamlets' rotate positively with longitude.  

Finally, the bugaboo question is whether these responses can be aliased, 
and the answer surely is yes, but it is also very unlikely that they are 
aliases of higher frequency responses.  Were the primary 7-$P_1$ 
modulation a first-order alias, its true frequency would be 0.86 c/$P_1$ 
and period correspondingly 1.16 $P_1$.  Such a modulation would have 
to be very very stable to correlate in the manner we find in Fig.~\ref{fig5}, 
and the 147-$P_1$ peak would be indicative of 126 beamlets!  We thus 
appear safe in proceeding on the assumption that the 7- and 147-$P_1$ 
features are not aliased.

\section*{VIII. Carousel Configuration} 
\label{sec8} 
We are now in a position to compute polar maps corresponding to the 
subpulse-modulation patterns discussed above.  A carousel circulation 
time $\P3hat$ of around 147 $P_1$ was indicated both directly by the 
low frequency feature and by ACFs of the PSs (Fig.~\ref{fig5}).  This, 
together with the 7-$P_1$ $P_3$ feature, strongly suggests a carousel 
with 20 beamlets.  We argued above that it is highly unlikely that the 
features are aliased, and longitude-longitude correlations appear to 
confirm that the carousel rotates positively with longitude through the 
outer conal components.   The longitude of the magnetic axis is taken 
at the midpoint between the outer conal component pair, the putative 
longitude origin used in most of the figures above.  And while we have 
not been able to determine $\alpha$ and $\beta$ more accurately, the 
``working'' values of 25 and +1.8\degr\ are sufficiently accurate for our 
present purposes---the latter implying an ``outside'' or poleward traverse 
of the sightline.  

\begin{figure}
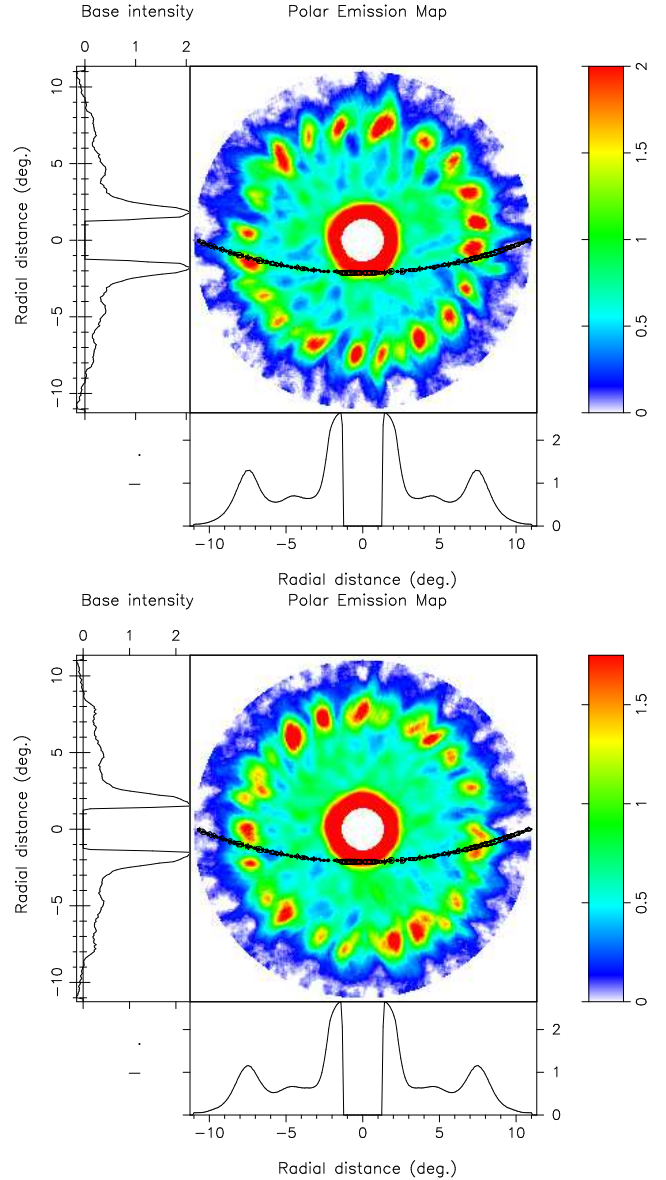
 
\begin{center}
\includegraphics[width=78mm,angle=-90.]{PQB1857-26_325_map_1101-1545_new.ps}
\includegraphics[width=78mm,angle=-90.]{PQB1857-26_325_map_1-588_new.ps}
\caption{ Polar maps constructed using the the respective pulse 1101-1545
(top) and 1-588 (bottom) PSs.  Note the roughly regular pattern of 20
`beamlets' in the top map and the irregular pattern in the lower one.
The magnetic axis is at the centre of the diagram and the ``closer''
rotational axis is upward, such that the sightline track sweeps through
the pattern as indicated.  Here the star rotates counterclockwise, causing
the sightline to cut the clockwise-rotating subbeam pattern from right to left;
see DR01 for further details.  The side panels give the ``base'' function
(which has not been subtracted from the maps here), and the lower
panels show the radial form of the average beam pattern.}
\label{fig7}
\end{center}
\end{figure}

Figure~\ref{fig7} displays carousel maps for two sections of the 325-MHz 
PS.  Both use a local value of $\P3hat$, computed as 20 times the $P_3$ 
determined within the same interval (7.42 and 7.21 $P_1$, respectively).  
The upper carousel map shows the average pattern of three rotations 
during a relatively stable interval (pulses 1101-1545; see Fig.~\ref{fig3}, 
top left); whereas the bottom map depicts the average configuration  
during a less ordered four-$\P3hat$ segment (\#1-588; see Fig.~\ref{fig3}, 
right).  There is a good deal to see in these diagrams.  First, there is  
substantial irregularity even in the upper map, such that 10 regularly 
spaced beamlets are only seen over half the map.  In the lower map, 
some beamlets have an azimuthal spacing near (360\degr/20=) 18\degr, 
but many do not, so that the broad primary modulation feature in the 
latter figure is not surprising.  Second, the diagrams show virtually no 
correspondence between the beamlet patterns in the inner and outer 
conal rings, as expected from our earlier lrf analyses.

\section*{IX.  Summary and Conclusions}
\label{sec9}
Five-component, core/double-cone pulsars are relatively rare in the 
normal pulsar population (\eg, Rankin 1993).  B1237+25 (\eg, Srostlik 
\& Rankin 2005) is by far the best studied example, and B1857--26 
has frequently been compared to it almost as a twin.  For the latter, 
however, little beyond average-profile studies have been available 
until very recently, and the present study provides the first 
sensitive PS analysis.  

Like B1237+25, B1857--26 exhibits both strong, regular subpulse 
modulation and null pulses.  In the latter, the $P_3$ is longer, about 
7.4 $P_1$ and is confined to the outer conal components.  Our two 
observations have so far identified only one behaviour, or profile 
mode, as against several in B1237+25.  The null fraction is also 
larger in B1857--26, about 20\%$\pm$3\%, and twice the value of 
10\%$\pm$2.5\% reported by Ritchings (1976).   And also like the 
former, the nulls in this pulsar are short, typically one pulse and no 
more than 4 pulses in our observations.  Possibly, these are in 
fact {\it pseudo} nulls [as in \eg,  B0834+06 (Rankin \& Wright 2007a) 
and B1133+16 (Herfindal \& Rankin 2007)] because the maximum 
null length is about $P_3/2$---and thus they represent ``empty'' 
sightline traverses through the subbeam carousel system.  If so, the 
4-$P_1$ nulls may occur preferentially in orderly or disorderly intervals 
of the observations.

Analyses of B1857--26 (unlike B1237+25) provide evidence of a 
tertiary modulation period $\P3hat$ of some 147 $P_1$ at both 325 and 610 MHz.  Indications 
of this tertiary period were first identified via an ACF analysis (\eg, 
Fig.~\ref{fig5}) and then corroborated by a low frequency lrf feature 
(Fig.~\ref{fig3}, bottom).  While the $\P3hat$ appears to vary in a 
range around 146 to 148 $P_1$ (just as measured values of $P_3$ 
vary between about 7.3-7.4 $P_1$), we have found that in any given 
interval the circulation time remains 20 times the primary modulation 
period.  Numerically, then, some 20 beamlets would then nominally 
comprise the carousel pattern, and the more or less regular (or irregular) intervals 
in our PSs appear to reflect differences in the constancy of spacing 
and possibly number of beamlets at a given time. 

The 147 $P_1$ circulation time is again very much larger than the 
value predicted by the R\&S (1975) theory.  For B1857--26, with a 
612-ms $P_1$ and a $B$ of 3$\times$10$^{11}$ G, this theory 
predicts only 5.6 $P_1$.  If ${\bf E}$$\times$${\bf B}$ forces drive 
the circulation, then it may be that they operate over a larger height 
range than envisioned in the above theory, as suggested, \eg, by 
Hibschman \& Arons (2001a,b) and Harding \etal\ (2002).  
Alternatively, a vaccum gap partially screened by thermionic ion 
flow in the inner acceleration region above the polar cap has been 
considered (Gil \etal\ 2003) to explain the observed slow drift 
rates in pulsars.

The PA traverse of B1857--26 presents an interesting case for 
further analysis.  If it is assumed that most of the star's emission 
stems from a single OPM, then the PA traverse exceeds 180\degr\ 
by an significant angle.  Alternatively, a different mode may 
predominate on the leading and trailing sides of the profile, 
or conceivably, the core emission may not follow the RVM.  The 
overall symmetry of the profile, however, does not necessarily 
support a change in the dominant OPM, as we know of no other 
good example of a pulsar with such a marked OPM asymmetry about the 
profile centre.  Such asymmetry is sometimes seen in profiles that 
represent an oblique sightline traverse, but we know of no well 
vetted example of such asymmetry where the sightline cuts the 
beam pattern centrally.  

Finally, the central core component of PSR B1857--26 provided us little basis 
for further analysis.  Its shape is not at all Gaussian, but otherwise 
it appears rather steady in both intensity and polarization, showing 
none of the effects exhibited by B0329+54 (Mitra \etal\ 2007).

\section*{Acknowledgments}
We thank Stephen Redman for assistance and the GMRT operational 
staff for observing support. We thank S. Sirothia and R. Athreya for 
insightful discussions. JMR sincerely thanks NCRA and its 
staff for their generous hospitality and support during a recent visit.   
This work made use of the NASA ADS system.

\bsp

\label{lastpage}


\begin{thebibliography}{99}
\bibitem[Asgekar \& Deshpande (2005)]{ad05} Asgekar, A., \& Deshpande, A. A. 2005, \mnras, 357, 1105
\bibitem[\protect\citeauthoryear{Backer}{1976}]{b76} Backer D. C. 1976, \apj, 209, 895
\bibitem[\protect\citeauthoryear{Barnard \& Arons}{1986}]{ba86} Barnard, J. J. \& Arons, J. 1986, 
	\apj, 302, 138
\bibitem[\protect\citeauthoryear{Blaskiewicz, Cordes \& Wassermann}{1991}]{bcw} Blaskiewcz, M., 
	Cordes, J. M., \& Wassermann, I. 1991 Ap.J., 370 643.
\bibitem[Deshpande \& Rankin (1999)]{dr99} Deshpande, A. A. \& Rankin,
          J.M. 1999, \apj, 524, 1008
\bibitem[Deshpande \& Rankin (2001)]{dr01} Deshpande, A. A. \& Rankin,
          J. M. 2001, \mnras, 322, 438
\bibitem[\protect\citeauthoryear{Everett \& Weisberg}{2001}]{n10}  Everett, J. E., \& Weisberg, J. M. 
	2001, \apj, 553, 341.
\bibitem[\protect\citeauthoryear{Gil \etal\ }{2003}]{n14}  Gil, J., Melikidze, G. I. \& Geppert, U., 
	2003, A\&A, 407, 315
	\mnras, 301, 253.
\bibitem[\protect\citeauthoryear{Gould \& Lyne}{1998}]{n15}  Gould, D. M., \& Lyne, A. G. 1998, 
	\mnras, 301, 253.
\bibitem[\protect\citeauthoryear{Gupta \etal\ }{2004}]{ggks}  Gupta, Y., Gil, J., Kijak, J., \& 
	Sendyk, M. 2004, \mnras, 426, 229.
\bibitem[\protect\citeauthoryear{Hamilton \etal\ }{1977}]{hmak} Hamilton, P. A., McCulloch, P. M., 	Ables, J. G., \& Komesaroff. M. M. 1977, \mnras,180, 1
\bibitem[\protect\citeauthoryear{Harding \etal\ }{2002}]{hmz02} Harding, A., Muslimov, A., \& 
	Zhang, B. 2002, \apj, 576, 366.  
\bibitem[\protect\citeauthoryear{Herfindal \& Rankin}{2007}]{HR07} Herfindal, J. L. \& 
	Rankin, J. M. 2007, \mnras, 380, 430
\bibitem[\protect\citeauthoryear{Hibschman \& Arons}{2001a}]{ha01a} Hibschman, J. A., 
	\& Arons, J., 2001a, \apj, 554, 624.  
\bibitem[\protect\citeauthoryear{Hibschman \& Arons}{2001b}]{ha01b} Hibschman, J.A., 
	\& Arons, J., 2001b, \apj, 560, 871.  
\bibitem[\protect\citeauthoryear{Hobbs \etal\ }{2005}]{hllk}   Hobbs, G.,  Lorimer, D. R., Lyne, A. G., 
	\& Kramer, M. 2005, \mnras, 360, 974
\bibitem[\protect\citeauthoryear{Johnston \etal\ }{2005}]{jhvkwl} Johnston, S., Hobbs, G., Vigeland, S., 
	Kramer, M., Weisberg, J. M., \& Lyne, A. G.  2005, \mnras, 364, 1397
\bibitem[\protect\citeauthoryear{Komesaroff}{1970}]{k70} Komesaroff, M. M. 1970, Nature, 225, 612
\bibitem[\protect\citeauthoryear{van Leeuwen \etal\ }{1988}]{lm88} van Leeuwen, A.G.J., 
	Stappers, B. W., Ramachandran, R., \& Rankin, J. M. 2003, \mnras, 399, 223
\bibitem[\protect\citeauthoryear{Lyne \& Manchester}{1988}]{lm88} Lyne, A. G., \& Manchester, R. N. 
	1988, \mnras, 234, 477
\bibitem[\protect\citeauthoryear{Manchester \etal\ }{1980}]{hmak} Manchester, R. N., Hamilton, P. A., \& 	McCulloch, P. M. 1980, \mnras,192, 153
\bibitem[\protect\citeauthoryear{McCulloch \etal\ }{1978}]{mhm} McCulloch, P. M., Hamilton, P. A., 	Manchester, R. N., \& Ables, J. G. 1978, \mnras,183, 645
\bibitem[\protect\citeauthoryear{Mitra \& Deshpande}{1999}]{md99} Mitra, D., \& 
	Deshpande, A. A. 1999, \aap, 346, 906
\bibitem[\protect\citeauthoryear{Mitra \etal\ }{2005}]{n32}  Mitra, D, Gupta, Y. \& Kudale, S., 
	2005, ``Polarization Calibration of the Phased Array Mode of the GMRT'', URSI GA 
	2005, Commission J03a
\bibitem[\protect\citeauthoryear{Mitra, Rankin \& Gupta}{2007}]{md99} Mitra, D., Rankin, J. M. \& 
	Gupta, Y. 2007, \mnras, 379, 932
\bibitem[\protect\citeauthoryear{Morris \etal\ }{1979}]{mgsbt} Morris, D., Graham, D. A., Sieber, W., 
	Jones, B. B., Seiradakis, J. H.,  \& Thomasson, P. 1979, \aap, 73, 46
\bibitem[\protect\citeauthoryear{van Ommen \etal\ }{1997}]{vOdAHM} van Ommen, T. D., 
	D'Alessandro, F., Hamilton, P. A., \& McCulloch, P. M. 1997, \mnras, 287, 307
\bibitem[\protect\citeauthoryear{Radhakrishnan \& Cooke}{1969}]{rc69} Radhakrishnan, V., 
	\& Cooke, D. J. 1969, Ap. Lett, 3, 225
\bibitem[\protect\citeauthoryear{Radhakrishnan \& Rankin}{1993}]{rr93} Radhakrishnan, V., 
	\& Rankin, J. M.. 1990, \apj, 352, 258.
\bibitem[\protect\citeauthoryear{Rankin}{1983}]{r83a} Rankin J. M. 1983a, \apj, 274 333
\bibitem[\protect\citeauthoryear{Rankin}{1983}]{r83b} Rankin J. M. 1983b, \apj, 274 359
\bibitem[\protect\citeauthoryear{Rankin}{1986}]{r86} Rankin J. M. 1986, \apj, 301, 901
\bibitem[\protect\citeauthoryear{Rankin}{1993}]{k7} Rankin J. M. 1993, \apj, 405, 285 and \apjs, 85, 145
\bibitem[\protect\citeauthoryear{Rankin}{2007}]{k8} Rankin J. M. 2007, \apj, 664, 443 (20 July 2007)

\bibitem[\protect\citeauthoryear{Rankin \& Ramachandran}{2003}]{rr03} Rankin J. M., Ramachandran R. 2003, Ap.J., 590, 411
\bibitem[Rankin \& Wright (2007a)]{rw07a} Rankin, J. M. \& Wright, G.A.E. 2007a, \mnras, 379, 507
\bibitem[Rankin \& Wright (2007b)]{rw07b} Rankin, J. M. \& Wright, G.A.E. 2007b, \mnras, submitted

\bibitem[\protect\citeauthoryear{Ritchings}{1976}]{r76} Ritchings, R.T. 1976, \mnras, 176, 249
\bibitem[Ruderman \& Sutherland 1975]{R75} Ruderman, M. A., Sutherland, P. G., 1975, \apj, 196, 51
\bibitem[Srostlik \& Rankin 2005]{sr05} Srostlik, Z., \& Rankin, J. M., 2005, \mnras, 362, 1121
\bibitem[\protect\citeauthoryear{Vaughan \& Large}{1970}]{k1} Vaughan, A. E. \& Large, M. I. 1970, \nat, 225, 167
\bibitem[Weltevrede \etal\ 2006]{wes06} Weltevrede, P., Edwards, R., \& Stappers, B., 2006, \aap, 445, 243.
\bibitem[Weltevrede \etal\ 2007]{wse07} Weltevrede, P., Stappers, B., \& Edwards, R., 2007, \aap, 469, 607.


\end{thebibliography}
\end{document}